\documentclass[pra,twocolumn,aps,epsfig,showpacs]{revtex4}
\usepackage{bm}
\usepackage{amsfonts}
\usepackage[dvips]{graphicx}
\usepackage{mathrsfs}
\usepackage[intlimits]{amsmath}
\begin{document}
\title{Proposed all-versus-nothing violation of local realism in the Kitaev spin-lattice model}
\author{Ming-Guang Hu}
\author{Dong-Ling Deng}
\author{Jing-Ling Chen}
 \email{chenjl@nankai.edu.cn}
\affiliation{Theoretical Physics Division, Chern Institute of
Mathematics, Nankai University, Tianjin 300071, People's Republic of
China}

\date{\today}


\begin{abstract}
We investigate the nonlocal property of the fractional statistics in
Kitaev¡¯s toric code model. To this end, we construct the
Greenberger-Horne-Zeilinger paradox which builds a direct conflict
between the statistics and local realism. It turns out that the
fractional statistics in the model is purely a quantum effect and
independent of any classical theory.We also discuss a feasible
experimental scheme using anyonic interferometry to test this
contradiction.
\end{abstract}

\pacs{03.65.Ud, 03.67.Mn, 75.10.Jm, 42.50.Dv}

\maketitle

Quantum theory can predict results which are never achievable from
the local realism (LR)~\cite{A.Einstein}. By definition, LR consists
of two constraints of realism and locality: Any observable has a
predetermined value, regardless of whether it is measured or not,
and the choice of which observable to measure on one party of a
multipartite system does not affect the results of the other
parties. These constraints lead not only to the well-known Bell
inequalities~\cite{J.S.Bell} which put bounds on the correlations
and are violated statistically by certain quantum states, but also
to the so-called Greenberger-Horne-Zeilinger (GHZ)
paradox~\cite{GHZ-TH} which derives directly the inconsistent values
of the correlations from LR and quantum mechanics. Such paradox is
tested by the nonstatistical measurements, yielding succinctly an
all-versus-nothing proof between LR and quantum mechanics.

In this work we investigate LR in the context of Kitaev's toric code
spin-lattice model, which is an exactly solvable model and is
crucial for fault-tolerant topological quantum computation
(TQC)~\cite{2003Kitaev,2006Kitaev}. This model has the merits: Its
degenerate ground states yield a topologically protected subspace
that provides robustness against noise and quasilocal perturbations,
arousing much interest in condensed matter and quantum optical
physics to realize and control it
\cite{2008Jiang,2006Micheli,2003Duan}; Excitations of the ground
states known as anyons possess a class of fractional statistics
\cite{1982Wilczek} intervening between the bosonic and fermionic
statistics, that is, the quantum state of anyons can acquire an
unusual phase factor when one anyon is exchanged with another one,
in contrast to usual values $+1$ for bosons and $-1$ for fermions.

Since anyons are at the heart of TQC \cite{2007Nayak}, it is natural
and important to ask whether the anyonic statistics, though defined
in quantum mechanics can be described by LR. With this aim, we
construct the GHZ paradox by using the string operators that are
used  in the model to move anyons on the lattice. According to this
paradox, the results derived from anyonic statistics will contradict
irreconcilably that derived from LR. In this way we conclude
straightforwardly that the fractional statistics in Kitaev's model
is purely a quantum effect and independent of any classical theory.
In experiment, the GHZ paradox was tested only by using multi-photon
systems \cite{2000Pan,1999Bouwmeester}. The model discussed here
also provides a potential platform to test the GHZ paradox in the
future and this is discussed briefly at the end.

\begin{figure}[tbph]
\includegraphics[width=8cm]{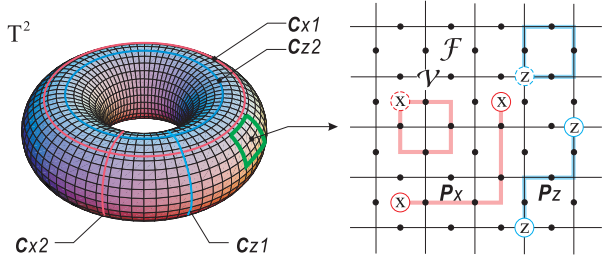}\\
\caption{(Color online) An illustration of the Kitaev spin-lattice
model:  Each black dot on the edge of the lattice represents a
spin-$1/2$ particle or qubit; The interactions of the Hamiltonian
$H_0$ are along edges that bound a face $\mathcal{F}$, and edges
that meet at a vertex $\mathcal{V}$. The string $P_{x,z}$ indicate
paths of products of $\sigma^{x,z}$ operators that are logical
operators on the qubits.}\label{fig1}
\end{figure}

The Kitaev's toric code spin-lattice model \cite{2003Kitaev} is
introduced as follows. Considering a $k\times k$ square lattice on a
torus ${\rm T}^2$ (see Fig. \ref{fig1}), one spin or qubit is
attached to each edge of the lattice. Thus there are $2k^2$ qubits.
For each vertex $\mathcal{V}$ and each face $\mathcal{F}$, consider
operators of the following forms:
\begin{equation*}
A_\mathcal{V}=\prod_{j\in \mathcal{V}}\sigma_j^x,\quad
B_\mathcal{F}=\prod_{j\in \mathcal{F}}\sigma_j^z,
\end{equation*}
where the $\sigma_j^X$ denotes the Pauli matrix with $X=x,y,z$ and
it acts on the $j$-th qubit of a vertex $\mathcal{V}$ or face
$\mathcal{F}$. These four-body interacting operators commute with
each other because a vertex $\mathcal{V}$ and a boundary
$\mathcal{F}$ consist of either $0$ or $2$ common qubits. From their
definitions, we know that operators $A_\mathcal{V}$ and
$B_\mathcal{F}$ have eigenvalues $\pm1$. Summing them together, it
constructs the model Hamiltonian as
\begin{equation*}
H_0=-\sum_{\mathcal{V}\in {\rm
T}^2}A_\mathcal{V}-\sum_{\mathcal{F}\in {\rm T}^2}B_\mathcal{F},
\label{eq-Hami}
\end{equation*}
of which the ground states $|{\rm g}\rangle$ satisfy
$A_\mathcal{V}|{\rm g}\rangle=|{\rm g}\rangle$,
$B_{\mathcal{F}}|{\rm g}\rangle=|{\rm g}\rangle$ for all
$\mathcal{V},\mathcal{F}$. Due to the topological property of torus,
the ground states are four-fold degenerate and construct a
four-dimensional Hilbert subspace, one basis of which can be
explicitly written as
\begin{equation}
|{\rm g_0}\rangle=\mathcal{J}\prod_{\mathcal{V}\in {\rm
T}^2}(1+A_\mathcal{V})|0\rangle^{\otimes 2k^2},
\end{equation}
with a normalization constant $\mathcal{J}$, while the remaining
three can be given after we introduce the concept of string
operators \cite{2005Hamma,2003Kitaev}.

Here we describe anyons as the quasiparticle excitations of the
spin-lattice system with $H_0$. There are two types of anyons:
$z$-particles living on the vertices and $x$-particles living on the
faces of the lattice. These anyons are created in pairs (of the same
type) by string operators: $|\psi^z(P_z)\rangle=S_{P_z}^z|{\rm
g}\rangle$ and $|\psi^x(P_x)\rangle=S_{P_x}^x|{\rm g}\rangle$ and
they live at the end of strings, where
\begin{equation}\label{eq-string}
S^z_{P_z}=\prod_{r\in P_z}\sigma_r^z,\quad S^x_{P_x}=\prod_{r\in
P_x}\sigma_r^x,
\end{equation}
are string operators associated with string $P_z$ on the lattice and
string $P_x$ on the dual lattice, respectively (Fig. \ref{fig1}).
Note that two anyons of the same type would annihilate  each other
when they meet and this is so called fusion rule. Then we can see
that $A_\mathcal{V}$ and $B_\mathcal{F}$ are just two closed string
operators. Especially, there are four nonequivalent classes of
closed strings that are not contractible, e.g.,
\{$\mathcal{C}_{x1}$, $\mathcal{C}_{z1}$, $\mathcal{C}_{x2}$,
$\mathcal{C}_{z2}$\} in Fig. \ref{fig1}. The corresponding string
operators \{$S^x_{\mathcal{C}_{x1}}$, $S^z_{\mathcal{C}_{z1}}$,
$S^x_{\mathcal{C}_{x2}}$, $S^z_{\mathcal{C}_{z2}}$\} have the same
commutation relations as \{$\sigma_1^x$, $\sigma_1^z$, $\sigma_2^x$,
$\sigma_2^z$\} and all of them commute with $A_\mathcal{V}$,
$B_\mathcal{F}$ and thus $H_0$, which consequently give out the
remaining three bases of the ground state subspace through
\{$S^x_{\mathcal{C}_{x1}}|{\rm g}_0\rangle$,
$S^x_{\mathcal{C}_{x2}}|{\rm g}_0\rangle$,
$S^x_{\mathcal{C}_{x1}}S^x_{\mathcal{C}_{x2}}|{\rm g}_0\rangle$\}
\cite{2005Hamma,2003Kitaev}. In addition, if one utilizes string
operators to move an $x$- (or $z$-) particle around a $z$- (or
$x$-)particle one loop, a global phase factor $-1$ would be picked
up in front of the initial wavefunction. This is the unusual
statistical property of abelian anyons.

\begin{figure}
\includegraphics[width=80mm]{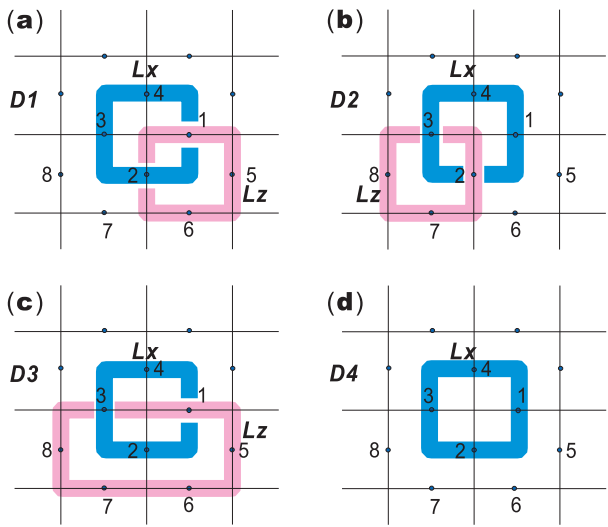}\\
\caption{(Color online) Using string operations to illustrate
Eqs.~(3). (\textbf{a}) The operator $D_1$ acts on a ground state
$|\rm g\rangle$; it means that first a pair of $z$-particles and
$x$-particles are created at the qubit labeled by $2$, then one of
the $x$-particles is moved along the loop $L_x$ and annihilate with
the other one, and finally the string operator
$S_{L_z^{651}}^z=\sigma_6^z\sigma_5^z\sigma_1^z$ moves one of the
$z$-particles to meet and hence annihilate another one along the
loop $L_z$. (\textbf{b}), (\textbf{c}) The operators $D_2$ and $D_3$
act on the ground state respectively and they have the similar
meanings as $D_1$. (\textbf{d}) The operator $D_4$ acts on the
ground state and contains only one loop $L_x$, which shows the
moving of $x$-particles. \label{fig2}}
\end{figure}

Now, we present our idea for the construction of GHZ paradox by
using the string operators. We define four composite string
operations denoted by $D_{1,2,3,4}$ by virtue of string operators
(Fig. \ref{fig2}). For $D_1$, it can be written as
$D_1=S_{L_z^{651}}^zS_{L_x}^xS_{L_z^2}^z$. When it acts on a ground
state $|{\rm g}\rangle$, anyons will be created, moved and
annihilated as follows:  First, a pair of $z$-particles are created
by the string operator $S_{L_z^2}^z=\sigma_2^z$ with $L_z^j$
denoting the edge where the $j$-th qubit crosses the loop $L_z$ as
shown in Fig. \ref{fig2}(a); Then a pair of $x$-particles are
created, moved and annihilated along the loop $L_x$ by the string
operator $S_{L_x}^x=\sigma_1^x\sigma_4^x\sigma_3^x\sigma_2^x$; At
last, the string operator
$S_{L_z^{651}}^z=\sigma_6^z\sigma_5^z\sigma_1^z$ moves one of the
$z$-particles to meet and hence annihilate another one. As a result,
we can see from Fig. \ref{fig2}(a) that the loops $L_x$ and $L_z$
construct a link. According to the anyonic statistics, a global
phase factor $-1$ is picked up in front of the ground state, i.e.,
$D_1|{\rm g}\rangle=-|{\rm g}\rangle$. Likewise, as for the
operations $D_2=S_{L_z^{872}}^zS_{L_x}^xS_{L_z^3}^z$, and
$D_3=S^z_{L_z^{87651}}S_{L_x}^xS^z_{L_z^{3}}$, they have the same
interpretations as $D_1$ but with different loops $L_{x,z}$. The
links constructed by $L_{x,z}$ for $D_{2,3}$ are shown in Fig.
\ref{fig2}(b) and \ref{fig2}(c), respectively and due to the anyonic
statistics, we have $D_{2,3}|{\rm g}\rangle=-|{\rm g}\rangle$. As
for the operation $D_4=S_{Lx}^x$, it has a straightforward
correspondence with $A_\mathcal{V}$ operator and its loop $L_x$
constructs a simple unknot shown in Fig. \ref{fig2}(d). Therefore
when acting it on the ground state, no change would appear, i.e.,
$D_4|{\rm g}\rangle=|{\rm g}\rangle$.

Writing the above statements of composite string operations into
more convenient forms, we have
\begin{subequations}
\begin{eqnarray}
D_1|{\rm g}\rangle&=&\sigma_6^z\sigma_5^z\sigma_4^x\sigma_3^x\sigma_2^y\sigma_1^y|{\rm g}\rangle=-|{\rm g}\rangle\label{D1},\\
D_2|{\rm g}\rangle&=&\sigma_8^z\sigma_7^z\sigma_4^x\sigma_3^y\sigma_2^y\sigma_1^x|{\rm g}\rangle=-|{\rm g}\rangle\label{D2},\\
D_3|{\rm g}\rangle&=&\sigma_8^z\sigma_7^z\sigma_6^z\sigma_5^z\sigma_4^x\sigma_3^y\sigma_2^x\sigma_1^y|{\rm g}\rangle=-|{\rm g}\rangle\label{D3},\\
D_4|{\rm g}\rangle&=&\sigma_4^x\sigma_3^x\sigma_2^x\sigma_1^x|{\rm
g}\rangle=|{\rm g}\rangle\label{D4},
\end{eqnarray}
\end{subequations}
in which the algebraic relation $\sigma_j^z\sigma_j^x=i\sigma_j^y$
has been used. In the viewpoint of measurements, suppose there are
eight observers and each of them has access to one spin (See Fig.
\ref{fig2}) in the model. On the $i$-th spin, the corresponding
observer measures the observable $\sigma_i^X$ without disturbing
other spins and the measurement result is denoted by $m^X_i$. Since
these results must satisfy the same functional relations satisfied
by the corresponding operator, then from Eqs. (3), we can predict
that, if all the operators in Eqs. (3) are measured, their results
must satisfy
\begin{subequations}
\begin{eqnarray}
m^z_6m^z_5m^x_4m^x_3m^y_2m^y_1&=&-1,\label{d1}\\
m^z_8m^z_7m^x_4m^y_3m^y_2m^x_1&=&-1,\label{d2}\\
m^z_8m^z_7m^z_6m^z_5m^x_4m^y_3m^x_2m^y_1&=&-1,\label{d3}\\
m^x_4m^x_3m^x_2m^x_1&=&+1.\label{d4}
\end{eqnarray}
\end{subequations}
After obtaining this, we next reveal how it produces the
contradiction according to LR.

Noting that Eqs.~(3) contain only local operators, the operators in
each equation thereby commute and can all simultaneously have their
eigenvalues.
Thus, from LR we can associate an element of reality to each of the
eigenvalues in Eqs.~(4). For instance, the observers on particles
($2, 3,4,5,6$) measure, without disturbing each other, the
observables
($\sigma_6^z,\sigma_5^z,\sigma_4^x,\sigma_3^x,\sigma_2^y$),
respectively and if the multiplier of their results is $1$ (or
$-1$), then from Eq.~(\ref{d1}) they can predict with certainty that
the result of measuring $\sigma_1^y$ will be $-1$ (or $1$). That is,
they can predict with certainty the value of quantity $\sigma_1^y$
by measuring other particles without disturbing particle $1$, and
therefore an element of reality can be associated to the physical
quantity $\sigma_1^y$. Analogously, we can associate elements of
reality to all the physical quantities in Eqs.~(3). Then we can
suppose that this result was somehow predetermined and initially
hidden in the original state of the system. Such predictions with
certainty would lead us to assign values $+1$ or $-1$ to all the
observables in Eqs.~(3). However, such assignment cannot be
consistent with rules of quantum mechanics because if we multiply
Eqs.~(\ref{d1})-(\ref{d3}) together, it will lead to,
$m^x_4m^x_3m^x_2m^x_1=-1$, which directly contradicts
Eq.~(\ref{d4}). Therefore, we conclude that the four predictions of
quantum mechanics given by Eqs.~(3) cannot be reproduced by LR. This
completes the construction of the GHZ paradox in the context of the
Kitaev spin-lattice model.

Further, the above GHZ paradox applies to more general situations.
We can enlarge the loops $L_{x,z}$ in Fig. \ref{fig2} to generalize
these $D_{1,2,3,4}$ operations. For each set of $D_{1,2,3,4}$ when
acting on a ground state, it can admit the GHZ paradox only if they
satisfy all of the following requirements: (i) the loop $L_x$ for
all of them should be the same, no matter how large area they
enclose; (ii) there is a loop $L_z$ for each of $D_{1,2,3}$ that
should construct a link when combined with $L_x$; (iii) when we
merge the $L_z$ of $D_1$ with the $L_z$ of $D_2$ together with
overlapping edges vanishing, the resultant loop should be the same
as the $L_z$ of $D_3$.  In this case, the string operators of anyons
give us a simple yet effective approach to look for various sets of
$D_{1,2,3,4}$ operators to construct the GHZ paradox.

To sum up, it turns out that the GHZ paradox is very common in the
Kitaev's toric code model. The all-versus-nothing violation of LR
above well shows the anyonic statistics in the model as a pure
quantum effect. In a way, it also indicates that the anyonic
statistics may be at the conflictive regime between LR and quantum
mechanics, which still needs an investigation in the future.

At the end, let us discuss briefly a feasible experimental
implementation of the above consideration. The Kitaev spin-lattice
model could be realized through dynamic laser manipulation of
trapped atoms \cite{2003Duan} or molecules \cite{2006Micheli} in an
optical lattice. In addition, an approach of anyonic interferometry
in atomic systems was as well suggested recently by Jiang et al.
\cite{2008Jiang} to measure topological degeneracy and anyonic
statistics, enabling the measurement of the statistical phase
associated with arbitrary braiding paths. By using this approach, it
suffices to implement the operations $D_{1,2,3,4}$ in Eqs. (3) and
to detect the sign. We briefly introduce this in the following.

Consider a spin lattice of trapped atoms or molecules inside an
optical cavity (as shown in Fig. 2a of Ref. \cite{2008Jiang}), which
provides a model Hamiltonian $H_0$ and on which the spins are called
memory qubits. Except for the memory qubits, an additional ancilla
spin is needed to probe the sign change before ground states and
hence is called the probe qubit. To achieve controlled-string
operations, an optical cavity associated with the quantum
nondemolition interaction between the common cavity mode and
selected spins is used to implement, e.g., a $z$-type string
operation
\begin{equation}
\Lambda[S_\mathcal{C}^z]=|1\rangle_{\rm A}\langle 1|\otimes
S_\mathcal{C}^z+|0\rangle_{\rm A}\langle0|\otimes {\rm I},
\end{equation}
where the probe qubit is spanned by $\{|0\rangle_{\rm
A},|1\rangle_{\rm A}\}$. It means: If the ancilla spin is in state
$|0\rangle_{\rm A}$, no operation is applied to the memory qubits;
If the ancilla spin is in state $|1\rangle_{\rm A}$, the operation
$S_\mathcal{C}^z$ is applied to the topological memory. For our
spin-lattice system with $H_0$, we prepare its initial state
$|\Psi_{\rm initial}\rangle$ to be a ground state $|{\rm g}\rangle$
and $D_j|\Psi_{\rm initial}\rangle=\pm|\Psi_{\rm initial}\rangle$.
Here the sign in front of the gound state $|{\rm g}\rangle$ is what
we need to observe. The following interference experiment can be
used to measure the sign. First, we prepare the probe qubit in a
superposition state $(|0\rangle_A+|1\rangle_A)/\sqrt{2}$. We then
use controlled-string operations to achieve interference of the
following two possible evolutions: If the probe qubit is in state
$|0\rangle_{\rm A}$, no operation is applied to the memory qubits;
If the probe qubit is in state $|1\rangle_{\rm A}$, the operation
$D_j$ is applied to the topological memory, which picks up the extra
phase factor $e^{i\theta_{j}}$ we want to measure. After the
controlled-string operations, the probe qubit will be in state
$(|0\rangle_{\rm A}+e^{i\theta_j}|1\rangle_{\rm A})/\sqrt{2}$.
Finally, we project the probe qubit to the basis of
$|\xi_\pm\rangle\equiv(|0\rangle_{\rm A}\pm e^{i\phi}|1\rangle_{\rm
A})/\sqrt{2}$ with $\phi\in[0,2\pi)$, and measure the operator
$\sigma_\phi\equiv|\xi_+\rangle\langle\xi_+|-|\xi_-\rangle\langle\xi_-|$.
The measurement of $\langle \sigma_\phi\rangle$ versus $\phi$ should
have fringes with perfect contrast and a maximum shifted by
$\phi=\theta_j$ for $\langle\sigma_\phi\rangle=\cos(\phi-\theta_j)$.
In other words, for sigma operations $D_j$ ($j=1,2,3$), the $\phi$
shifts of the maximal $\langle\sigma_\phi\rangle$ will differ from
those for $D_4$ by $\pi$.

In summary, we have shown the GHZ paradox in the context of Kitaev's
toric code spin-lattice model by using the anyonic string
operations. It shows that the anyonic statistics in the model cannot
be described by LR but be a purely quantum effect. In return, the
Kitaev model provides a potential platform for testing the GHZ
paradox or LR in the future. A feasible experimental consideration
by using the anyonic interferometry is discussed to test such a
contradiction at the end. It is worth noting that the measurement
employed in the above experimental scheme is non-destructive and can
be repeated without disturbing the ground state. It is a predominant
advantage of using Kitaev's model compared with the experimental
tests by using multi-photon systems. Also recent experiments
demonstrated string operations on small networks of interacting NMR
qubits~\cite{2007Du} and non-interacting optical
qubits~\cite{2007Pachos,2007Lu}, on the basis of which it is
possible to realize our construction in advance on a small-scale
qubit system.

Besides, the ground states of the Kitaev model belong to graph
states, which is crucial in quantum information application. Bell
inequalities have been shown to discuss LR for graph states
\cite{2005Guhne}. Our construction here actually also contributes to
the subject.

M. G. H. is indebted to J. Du for enlightenment at USTC. This work
was supported in part by NSF of China (Grants No. 10575053 and No.
10605013), Program for New Century Excellent Talents in University,
the Project-sponsored by SRF for ROCS, SEM, and LuiHui Center for
Applied Mathematics through the joint project of Nankai and Tianjin
Universities.

\end{document}